# Fate of gadolinium in inflamed mouse brain: Release and phosphate interaction post-contrast agent administration

**Short title:** Gadolinium retention post-contrast injection

**Authors:**

Lina Anderhalten, MD/PhD[1,*], Nicole Höfer, MSc[2,*], Daria Dymnikova, MSc[2,*], Julia Hahndorf, PhD[3], Matthias Taupitz, MD[3], Heike Traub, PhD[4], Christian Teutloff, PhD[2], Carmen Infante-Duarte, PhD[1], Robert Bittl, PhD[2]

**Affiliations:**

[1] Experimental and Clinical Research Center, A Cooperation Between the Max Delbrück Center for Molecular

   Medicine in the Helmholtz Association and Charité - Universitätsmedizin Berlin, 13125 Berlin, Germany.

[2] Freie Universität Berlin, Fachbereich Physik, 14195 Berlin, Germany.

[3] Charité - Universitätsmedizin Berlin, Corporate Member of Freie Universität Berlin and Humboldt-Universität zu Berlin, Department of Radiology, Charitéplatz 1, 10117 Berlin, Germany.

[4] Bundesanstalt für Materialforschung und -prüfung (BAM), Division Inorganic Trace Analysis, 12489 Berlin, Germany.

* equally contributing authors

**Correspondence to:** Prof. Dr. Carmen Infante-Duarte, Experimental and Clinical Research Center (ECRC), Charité–Universitätsmedizin Berlin, Campus Buch, Lindenberger Weg 80, 13125 Berlin, Germany. Phone: +4930450539055, E-mail: carmen.infante@charite.de. Prof. Dr. Robert Bittl, Freie Universität Berlin, Fachbereich Physik, Arnimallee 14, 14195 Berlin, Germany. Phone: +49 30 838 56049, E-Mail: robert.bittl@fu-berlin.de. ORCID: 0000-0003-4103-3768.

**Supplemental digital contents are available for this article.**

**Author Contributions:** Conceptualization of the project: CID, CT, and RB; performing in vivo mouse experiments and ex vivo hippocampal organotypic brain slice cultures: LA; tissue processing and EPR sample preparation: LA, NH, and DD; performing EPR/ENDOR experiments: NH and DD; EPR data evaluation and analysis: NH, DD, CT, and RB; sample preparation for MRI calibration, MRI acquisition and formal data analysis: LA, JH, and MT; LA-ICP-MS data acquisition and interpretation: LA and HT; data visualization: LA, NH, and DD; writing - original draft preparation: LA, NH, and DD; writing – editing and final manuscript: LA, NH, CID, and RB with contributions from all coauthors; funding acquisition and supervision: MT, HT, CT, CID, and RB. All authors have read and agreed to the submitted version of the manuscript.

**Conflicts of interest:** There are no conflicts of interest to report.

**Sources of funding:** We gratefully acknowledge financial support from the Hertie Foundation to L. Anderhalten (medMS program) and funding by Deutsche Forschungsgemeinschaft (DFG) within the Collaborative Research Center (CRC) 1340, "Matrix in Vision" (projects A03, B03, B05, and C02), Grant No. 372486779.

# Fate of gadolinium in inflamed mouse brain: Release and phosphate interaction post-contrast agent administration

**ABSTRACT**

**Objectives:** In view of the mounting evidence for markedly increased cerebellar Gd retention under neuroinflammatory conditions after repeated linear GBCA administration *in vivo*, we aimed to discriminate between Gd retained within the GBCA complex and forms dissociated form the complex within the CNS and to characterize the chemical environment of the released $Gd^{3+}$. For this purpose, we employed electron paramagnetic resonance (EPR) spectroscopy, which enables direct detection of $Gd^{3+}$ release and evaluation of its molecular surroundings in intact cerebellar tissue from inflamed and non-inflamed brain sections exposed to either linear or macrocyclic GBCAs.

**Materials and Methods:** We performed EPR and electron nuclear double resonance (ENDOR) experiments on sub-mm brain samples taken after administration of linear gadopentetate and macrocyclic gadobutrol *in vivo* in a murine multiple sclerosis model and *ex vivo* in organotypic hippocampal slice cultures under inflammatory conditions. Complementary mass spectrometry (MS) and MRI calibration experiments on Gd-spiked homogenized mouse brain tissue were used to identify potential Gd bindings and assess the relaxation-active fraction of retained Gd.

**Results:** EPR detected µM range Gd levels in intact cerebellar biopsies and slices and distinguished between complex-bound and released Gd following linear GBCA administration *in vivo*. In biopsies, we detected by ENDOR phosphorus-containing molecules in the microenvironment of released Gd. Binding to inorganic ligands was evidenced using MS and MRI calibration experiments in homogenized mouse brain tissue.

**Conclusions:** EPR and ENDOR proved to be sensitive methods for detecting $Gd^{3+}$ release and characterizing retained Gd species within intact brain tissue. Our findings demonstrate inflammation-promoted Gd retention, underscore the importance of integrating *in vivo* and *ex vivo* analyses to unravel mechanisms of long-term Gd retention, and suggest that conventional

MRI may underestimate the true extent of Gd accumulation, especially under neuroinflammatory conditions.

**Keywords:**

electron paramagnetic resonance; experimental autoimmune encephalomyelitis; gadolinium-based contrast agents; gadolinium retention; laser ablation inductively coupled plasma mass spectrometry; magnetic resonance imaging; multiple sclerosis; neuroinflammation, organotypic hippocampal slice cultures, phosphate interaction

## 1. INTRODUCTION

Gadolinium (Gd) retention following the administration of gadolinium-based contrast agents (GBCAs) for magnetic resonance imaging (MRI) has obtained considerable scientific interest over the past decade. Repeated intravenous (i.v.) injections of GBCAs based on linear Gd complexes (linear GBCAs) appear to result in Gd retention in different human body tissues (1, 2). Since free ionic Gd ($Gd^{3+}$) is known to be highly cytotoxic, multiple studies were performed to investigate the mechanisms of Gd retention (1). For the central nervous system (CNS), this phenomenon was reported for the first time in 2014 and showed that the cerebellar nuclei (CN) are sites prone to Gd retention (3). Subsequent investigations, comprising numerous human (4-8) and animal studies (9-14), have consistently validated sustained Gd retention within the brain *in vivo* after linear GBCA administration. Contrary to that, Gd retention for GBCAs containing macrocyclic Gd complexes (macrocyclic GBCAs) was observed to be only transient, following compound-specific washout kinetics (15-19). As a result, the European Medicines Agency restricted the clinical utilization of linear GBCAs (20), while the US Food and Drug Administration released class warnings and recommendations to limit their use (21).

The probability of Gd deposition in brain tissue was reported to depend on the GBCA-specific kinetic and thermodynamic stability, which is higher for macrocyclic compared to linear GBCAs (2, 22, 23). While macrocyclic Gd complexes are expected to be cleared from the brain as intact chelates (24-27), linear Gd complexes may gradually dissociate *in vivo* and release

Gd$^{3+}$ from their chelating ligand when entering brain areas rich in competing endogenous ions such as calcium (Ca) (28), copper (Cu) (29, 30), zinc (Zn) (28-30), and iron (Fe) (28-31). Free Gd$^{3+}$ may react with anionic inorganic phosphates and precipitate, forming long-lasting insoluble deposits of Gd phosphate salt (GdPO$_4$) within the CNS (22, 32). In addition, Gd$^{3+}$ may also bind to endogenous macromolecules with complexing abilities (11, 24, 25, 33), like adenosine triphosphate (ATP) (34). Recent studies have identified potential macromolecular Gd$^{3+}$-binding partners, such as ferritin (32), citrate (35), or glycosaminoglycan structures found in the extracellular matrix (ECM) or on cell surfaces (33, 35, 36). However, the nature of these deposits remains unclear.

In patients with the chronic neuroinflammatory disease multiple sclerosis (MS), Gd retention in the brain was shown using MRI after multiple injections of linear GBCAs (4, 37-41) and macrocyclic GBCAs (42, 43), yet the latter was subject to controversy (44, 45). In experimental autoimmune encephalomyelitis (EAE), a common mouse model of MS, we previously demonstrated that brain inflammation can markedly increase cerebellar Gd retention levels after repeated i.v. injections of the linear GBCA gadopentetate (46, 47) and alter its spatial distribution patterns, even over extended periods (47). In contrast, after application of the macrocyclic GBCA gadobutrol, Gd retention was only initially promoted by EAE inflammation to a comparable relative extent but was efficiently cleared from the brain over time (47). Thus, elucidation of the causes that trigger Gd$^{3+}$ release *in vivo* and facilitate sustained and potentially permanent Gd retention within the CNS is needed, such as a proinflammatory milieu and/or endogenous tissue factors with destabilizing or transchelating properties.

The quantification of Gd retention within CNS tissue following GBCA administration *in vivo* has been mainly conducted by techniques such as inductively coupled plasma mass spectrometry (ICP-MS) or laser ablation ICP-MS (LA-ICP-MS) (29, 47-50). To be able to distinguish between free Gd$^{3+}$, intact GBCA, and deposited Gd species, ICP-MS can be combined with chromatography techniques, which, however, require complex sample processing, potentially influencing the local Gd environment (24, 25, 51). Electron paramagnetic resonance (EPR) techniques have the potential for detection of $Gd^{3+}$ release and the determination of its local molecular environment without sample manipulation. However, while EPR is widely used in *in cell* studies, especially to explore protein conformations (52, 53), its utilization to study intact tissue samples is scarce.

In this study, we utilized EPR and electron nuclear double resonance (ENDOR) spectroscopy to analyze Gd retention and its chemical environment in intact cerebellar biopsies of EAE mice and a healthy control (HC) mouse following repeated *in vivo* administration of the linear GBCA gadopentetate or the macrocyclic GBCA gadobutrol. To assess the stability of Gd retention and the Gd environment, we also conducted EPR measurements on homogenates of chronic organotypic hippocampal slice cultures exposed to GBCA *ex vivo* under inflammatory conditions. Since magnetic resonance-based techniques like *in vivo* MRI and EPR may be limited in their ability to detect inorganic precipitated forms of Gd, we additionally performed an MRI *in vitro* calibration to determine the relaxation active amount of retained Gd and thereby estimate inorganic Gd bindings.

## 2. MATERIALS AND METHODS

### 2.1 Animal Ethics

Mouse experiments were performed in accordance with the European Communities Council Directive of 22 September 2010 (2010/63/EU). All experimental procedures were approved by the Berlin State Office for Health and Social Affairs (LAGeSo; approval-IDs: G106/19, TCH0008/20).

### 2.2 Mouse model of EAE and *in vivo* administration of GBCAs

EAE was actively induced in 10-week-old female SJL/J mice (Janvier Labs, France, n=2) by subcutaneous immunization with proteolipid protein peptide ($PLP_{139-151}$; purity 95%; Pepceuticals, Leicester, United Kingdom) (46, 47). Starting at the peak of disability (day 12 post-immunization), EAE mice were exposed to repeated i.v. injections of either a linear GBCA (*n*=1; gadopentetate dimeglumine, Magnevist, 0.5 mmol/ml, Bayer, Germany) or macrocyclic GBCA (*n*=1; gadobutrol, Gadovist, 1.0 mmol/ml, Bayer, Germany) at a cumulative dose of 20 mmol/kg body weight (BW). As illustrated in Figure 1a, four consecutive daily GBCA injections were followed by a two-day break and four additional injections (overall 8 x 2.5 mmol/kg BW), as previously described (46, 47). One HC mouse received gadobutrol following the same injection regimen. 10 days after the last administration of GBCAs, mice were sacrificed with an overdose of Ketamine/Xylazine (50 mg/ml Ketamine, 20 mg/ml Xylazine, in 0.9% NaCl). An untreated HC mouse of the same strain, age, and sex was used as a negative control. Mouse bodies were transcardially perfused with 50 ml of 0.1 M PBS (Ismatec ISM444B-115 V Analog Peristaltic Pump). Subsequently, brains were extracted and fixed in 4% paraformaldehyde (PFA) and 30% sucrose. Thereafter, they were coronally separated into the forebrain, midbrain, and cerebellum and stored at –80°C.

2.3 Collection of biopsies from mouse cerebellum

Cerebellar tissue from EAE mice ($n$ = 2) treated with gadopentetate or gadobutrol, from an HC mouse ($n$ = 1) treated with gadobutrol, and from an untreated HC mouse ($n$ = 1, negative control) was used for the evaluation of Gd retention and possible Gd complex dissociation *in vivo*. Mouse cerebella were thawed at room temperature, washed in 1 M sodium chloride, and transferred to a clean glass slide with the caudal part facing down. Three distinct brain regions were biopsied from each cerebellum, resulting in a total of three biopsy samples per mouse brain. The cerebellar regions of interest (ROI) were chosen based on our recently published data on Gd retention patterns in EAE brains using LA-ICP-MS (47). As displayed in Figure_1b, biopsies were collected from i) the cerebellar cortex, ii) the CN, and iii) the medulla using quartz capillaries for EPR (Vitrocom, Mountain Lakes, USA) with 0.87 mm/0.7 mm outer/inner diameters (OD/ID). To obtain sufficient tissue material in the EPR capillaries, each brain ROI was sampled four times (Figure 1b). Biopsy-containing capillaries were then centrifuged at 100 - 200 g for 2 min (Eppendorf Centrifuge 5417c, Eppendorf AG, Germany) to ensure that the tissue was positioned at the closed bottom of the capillaries for subsequent EPR measurements.

## 2.4 Generation of organotypic hippocampal tissue slice cultures

Chronic organotypic brain slices of the hippocampus were generated following an established protocol (54). Briefly, 7- to 10-day-old pups of the B6.Cg-Tg(Thy1-CFP/COX8A)S2Lich/J strain (FEM animal facility, Berlin Buch, Germany) were sacrificed by decapitation. The scalp was removed and the skull was placed on chilled cutting medium (1× MEM with 1% L-Glutamine). Under sterile conditions, the skull was opened, and the hippocampi were isolated and sliced into 350 µm-thick sections using a McIlwain Tissue Chopper (Mickle Laboratory Engineering Co., Ltd., Guildford, Surrey, UK). Intact slices were placed onto polytetrafluoroethylene (PTFE) membrane culture inserts (PICM0RG50, Millicell cell culture inserts, 0.4 µm) in 6-well plates in a randomized manner (6-8 slices per membrane, 3 replicate membranes per culture condition). Slices were cultured at 37°C and 5% $CO_2$ using a modified culture medium based on the study by Wang et al. (55) (1× MEM, 25% HBSS, 25% heat-inactivated horse serum, 13 mM HEPES, Pen/Strep (ingredients from Gibco™, Thermo Fisher Scientific Inc., Waltham, MA, USA), and 35 mM glucose (B. Braun, Melsungen, Germany)), which was changed every second day.

## 2.5 Chronic tissue slice culture treatment with GBCAs *ex vivo*

After 13 days in culture, slices were exposed to either gadopentetate or gadobutrol at 1 mM for 48 h, in line with our previous study (47) and as illustrated in Figure 2. To mimic an inflammatory milieu, hippocampal slices were additionally exposed to tumor necrosis factor α (TNFα; mouse recombinant, lyophilized; Invitrogen) at 50 ng/ml. Slices treated with TNFα without GBCA served as negative controls. For optimal substance penetration into the hippocampal tissue, 100 µl of the respective substance-enriched medium was applied on top of the membranes with the slices (3 conditions: *1*) TNFα; *2*) 1 mM gadopentetate + TNFα; *3*) 1 mM gadobutrol + TNFα). On day 15, to stop treatment, the membranes were washed repeatedly with fresh pre-warmed culture medium (3x10 min at 37°C, 5% $CO_2$; see Figure 2, To ensure a precise 48 h incubation per condition, treatment initiation on day 13 was time-staggered to match termination 48 h later on day 15. Thereafter, slices were incubated for one more week without additives; culture medium was changed every two days (see Figure 2).

## 2.6 Processing of chronic organotypic hippocampal tissue slice cultures

On day 23, membranes were cut out from the respective cell culture inserts with a sterile scalpel (*n*=3 per condition) and transferred into a 2 ml Eppendorf tube containing 0.5 ml of fresh culture medium. Using a mechanical homogenizer with a sterile pestle (Argos Technologies, Inc., Illinois, USA), the tube content was homogenized for 2 min to detach tissue samples from the PTFE membranes. Membranes were discarded, the tubes centrifuged for 2 min (400 g, Eppendorf Centrifuge 5417R, Eppendorf AG, Germany), and the supernatant was removed. The remaining tissue pellet was further washed with fresh culture medium in two more rounds of homogenization and centrifugation to remove unbound GBCA from the tissue sample. In the last step, the tissue pellet was homogenized with 20 µl medium for 30 seconds. The described process was repeated for each culture condition. As shown in Figure 2, the resulting tissue homogenates were stored at –80°C before they were thawed and filled into quartz capillaries (Vitrocom, Mountain Lakes, USA) with 0.87 mm/0.7 mm OD/ID, centrifuged and frozen for EPR assessment.

## 2.7 Preparation of EPR standards

EPR spectroscopy exclusively detects paramagnetic species, specifically ionic $Gd^{3+}$ and $Mn^{2+}$. For clarity, only the aqueous or released Gd ion is referred to as $Gd^{3+}$ throughout the text, while hyperfine couplings are denoted as Gd-H, Mn-H, Gd-P, and Mn-P, respectively. Similarly, the terms 'Gd complex', 'Gd environment' and 'retained Gd' also refer to $Gd^{3+}$ ions, especially in the context of EPR measurements. Since Mn can exist in various oxidation states, a distinction is made between $Mn^{2+}$ detected by EPR and Mn as the total manganese content assessed by LA-ICP-MS measurements.

GBCA standards for EPR measurements were prepared from gadopentetate dimeglumine (Magnevist, 0.5 mmol/ml, Bayer, Germany) and gadobutrol (Gadovist, 1.0 mmol/ml, Bayer,

Germany) by dilution in purified water and glycerol (1:1) to a final concentration of 0.03 mM. Several samples were prepared as references for $Gd^{3+}$-phosphate interactions i) Gd-ATP – aqueous solutions of $GdCl_3$ (gadolinium chloride hexahydrate, Sigma Aldrich) and ATP (adenosine 5'-triphosphate disodium salt trihydrate, Sigma Aldrich) were mixed with purified water and 50% glycerol to result in final concentrations of 0.3 mM $GdCl_3$ and 0.6 mM ATP; ii) GBCA-ATP – in analogy to Gd-ATP; iii) Gd-DMPC – an aqueous solution of DMPC (1,2-dimyristoyl-sn-glycero-3-phosphocholine, Avanti Polar Lipids, Inc.) vesicles was produced by first dissolving the DMPC powder in chloroform and then evaporating the chloroform in a low-pressure chamber. The resulting dry film of DMPC was dissolved in purified water. To decrease the proportion of double-walled vesicles, several freeze-thawing cycles and sonication were performed. An aqueous solution of $GdCl_3$ and 50 % glycerol was added to result in final concentrations of 0.1 mM $GdCl_3$ and 10 mM DMPC.

All samples were contained within quartz capillaries (Vitrocom, Mountain Lakes, USA) with 0.87 mm/0.7 mm OD/ID for EPR measurements, frozen and stored at –80 °C.

## 2.8 EPR

EPR measurements were performed on pure gadopentetate and gadobutrol, on the brain biopsy samples from the *in vivo* mouse experiment (4 conditions: 1. untreated HC mouse (= negative control), 2. gadopentetate-treated EAE, 3. gadobutrol-treated EAE, 4. gadobutrol-treated HC), on the tissue homogenates of the chronic hippocampal slice cultures referred to as '*slices*' (3 conditions: *1.* TNFα (= negative control), *2.* 1 mM gadopentetate + TNFα, *3.* 1 mM gadobutrol + TNFα), and on the reference samples Gd-ATP, gadopentetate-ATP, gadobutrol-ATP, and Gd-DMPC. For the measurements, the highest microwave frequency available, W-band (94 GHz), was used due to the high sensitivity (~ 1 µM) and small sample volumes required (~ 1 µl). W-band offers the additional advantage of spectra simplification for systems with high electron spin S, which is true for both $Gd^{3+}$ (S = 7/2) and unavoidable $Mn^{2+}$ (S = 5/2) components. The central transition ($m_S$ = –1/2 ↔ $m_S$ = +1/2) is broadened by the zero-field splitting only to second order, while the higher $m_S$ transitions are broadened to first order. This leads to a larger relative signal intensity of the central transition at higher magnetic fields/microwave frequencies (56-58). As a result, the spectra presented in this study primarily represent the central transition, albeit with some contribution from the broader component originating from higher $m_S$ transitions.

Continuous wave (cw)EPR and pulsed ENDOR were performed on a Bruker Elexsys E680 W-band (94 GHz) EPR spectrometer using a Bruker Teraflex EN600-1021H probe head. For ENDOR experiments, a Bruker DICE-II unit operating in stochastic mode and a radiofrequency (RF) amplifier AR-150A400 (Amplifier Research, Souderton, USA) were additionally used.

All cwEPR spectra were recorded with lock-in detection (100 kHz field modulation frequency with an amplitude of 0.5 mT and a conversion time of 60 ms), 80 µT field steps, and at a temperature of 80 K. As usual, the cwEPR spectra are presented as the first derivative, while those from pulsed experiments as absorption spectra.

ENDOR measurements give hyperfine couplings (peak splittings in the spectrum) and thus the distances between $Gd^{3+}$ or $Mn^{2+}$ and close-by nuclei with magnetic moments, such as hydrogen ($^1H$) and phosphorus ($^{31}P$). For the ENDOR experiments, separation of the $Gd^{3+}$ and the $Mn^{2+}$ signal contributions was required to distinguish between nuclei bound to $Gd^{3+}$ and $Mn^{2+}$, respectively. This was done by employing a filtering method utilizing the different transition moments of $Mn^{2+}$ and $Gd^{3+}$ (based on (59)). The following sequence, based on the Mims ENDOR technique (60), was applied to suppress the $Mn^{2+}$ signal and to measure Gd-H couplings: $\pi/2_{Gd}$ (18 ns) – 200 ns – $\pi/2_{Gd}$ (18 ns) – 1 µs – RF pulse (24 µs) – 2 µs – $\pi_{Mn}$ (48 ns). The "Gd" index indicates that the pulse flip angle was optimized for $Gd^{3+}$, e. g. $\pi_{Gd}$ is a π-pulse for $Gd^{3+}$. The microwave power was optimized so that pulse lengths of $\pi_{Gd}$ = 36 ns and $\pi_{Mn}$ = 48 ns were obtained. A frequency resolution of at least 16 kHz for $^1H$ ENDOR and 10 kHz for $^{31}P$ ENDOR measurements was used. For the measurements on the slices incubated with gadobutrol, signal separation was unnecessary due to a strong $Gd^{3+}$ signal and negligible $Mn^{2+}$ contribution. Thus, the sequence was adjusted to the standard $\pi/2_{Gd}$ (18 ns) – 200 ns - $\pi/2_{Gd}$ (18 ns) – 1 µs – RF pulse (24 µs) – 2 µs – $\pi/2_{Gd}$ (18 ns) sequence.

For Gd-P hyperfine couplings, the same sequences were used but with an initial delay between the $\pi/2_{Gd}$-pulses of τ = 470 ns; the same RF pulse length was maintained with the RF power increased by 6 dB in order to compensate for the lower nuclear magnetic moment of $^{31}$P.

$Mn^{2+}$-H couplings were recorded on the lowest-field $Mn^{2+}$ peak, where only a negligible $Gd^{3+}$ signal was present. Therefore, no separation was needed, and the following sequence was used: $\pi/2_{Mn}$ (24 ns) – 200 ns - $\pi/2_{Mn}$ (24 ns) – 1 µs – RF pulse (24 µs) – 2 µs – $\pi/2_{Mn}$ (24 ns). All ENDOR spectra were recorded at a temperature of 10 K.

2.9 Analysis of EPR and ENDOR spectra

All cwEPR spectra were analyzed and plotted using Matlab version 9.10 (61). A linear background was subtracted, and all spectra recorded on the biopsies after GBCA-treatment were normalized to the peak-to-peak amplitude of the third $Mn^{2+}$ line on the low field side, as were the spectra of the slices. Then, the spectrum of the respective control sample without GBCA treatment was subtracted, scaled individually to minimize the $Mn^{2+}$ signal in each difference spectrum (see Figures SDC_1 and SDC_2, Supplemental Digital Content 1, which visualize cwEPR difference spectra). The normalization on $Mn^{2+}$ is justified, as previous cerebellar LA-ICP-MS data on Mn showed a relatively homogeneous distribution of Mn across the different brain sections studied, eliminating the uncertainty in the sample amount mentioned above. These reference Mn data were obtained in the context of our previous study but were not included in Anderhalten *et al.* (47)). A lower limit for the concentration of retained Gd can then be achieved under the assumption of an optimal filling of the capillary with the largest $Mn^{2+}$ signal by comparing the signal intensity of interest with the GBCA reference spectrum.

To facilitate the identification of the peak positions in the ENDOR spectra, the $^1$H ENDOR spectrum of the gadopentetate-treated EAE mouse biopsy, as well as the $^{31}$P ENDOR spectrum of the Gd-DMPC reference spectrum, were symmetrized around the central frequency. Spectra of low signal-to-noise ratio were smoothed by using a binomially weighted moving average taken over a window of maximally 5 (cwEPR), 11 ($^1$H ENDOR), and 17 ($^{31}$P ENDOR) points. As further shown in Figure SDC_3 (see Supplemental Digital Content 2, including details on P ENDOR simulation), Gd-P distances were extracted by fitting the ENDOR spectra using the *saffron* and *esfit* functions of the Easyspin package (62).

## 2.10 MRI calibration in homogenized mouse tissue

Determination of local Gd concentrations in tissue by *in vivo* $T_1$-mapping is hampered by the fact that the paramagnetic efficacy ($^1$H NMR $T_1$ relaxivity ($r_1$) (63)) of Gd depends on various factors, mainly the molecular tumbling rate/molecular weight of the Gd-containing molecule, chemical environment, temperature, viscosity and others (35, 64). Vendor provided nominal $r_1$ of the GBCA cannot be applied directly. The comparison of Gd concentrations from *in vivo* $T_1$-mapping and other quantitative analytical methods requires knowledge about the concentration dependent $r_1$ of the GBCA in the tissue under the specific MRI conditions.

Therefore, an *in vitro* MRI calibration was conducted using brain tissue from eighteen 10-week-old female healthy SJL/J mice. Brains were isolated, flash-frozen in liquid nitrogen, and preserved at –80 °C. Thereafter, the brains were cryo-pulverized using the CP02 cryoPREP Automated Dry Pulverizer (Covaris Ltd, UK) and manually homogenized with 5% fetal bovine serum (FBS). To obtain GBCA standards ranging from 20 to 100 µM (in 20 µM steps), either gadopentetate or gadobutrol was added to the homogenate at varying concentrations (11). For baseline acquisition, samples without GBCA were prepared. Samples were transferred into 5 mm glass tubes and incubated for 30 minutes at 4 °C. Subsequently, the tubes were warmed up to 37 °C and MRI scans were performed on a 7 T small-animal scanner (Bruker BioSpec, Ettlingen, Germany), running ParaVision 6.1 software. An axial 2-dimensional $T_1$ map RARE-VTR sequence in analogy to (47) was used (echo time = 9.83 ms, 8 TRs from 255 to 7000 ms, rare factor = 2, field of view = 26 mm$^2$, matrix = 128 x 128, number of slices = 3, slice thickness: 1 mm, scan time: 17 min and 13 s). For internal validation, *in vitro* MRI calibration was also performed in water and pure FBS following the same protocol (36).

First, we quantified $r_1$ of both GBCAs to induce $T_1$ relaxation time shortening in homogenized mouse brain tissue. In brief, $T_1$ *in vitro* was computed by applying rotund ROIs to the tube regions of each T1 map RARE-VTR image, and values from the two lower MRI slices per condition were averaged (see Figure SDC_4a, Supplemental Digital Content 3). Linear regression analysis was performed using R software (RStudio, version 2022.07.2) by graphing $R_1$ at the different gadopentetate and gadobutrol concentrations, with the slope of $R_1$ regression lines representing $r_1$ ($r_1 = \Delta R_1/\Delta C_{Gd}$) (33, 36).

Second, Gd concentrations in the medulla, cerebellar cortex, and CN regions were quantified using LA-ICP-MS on cerebellar cryosections, as previously acquired and described in (47). While that study reported Gd levels in the CN region, the quantification of Gd concentrations for medulla and cortex regions 10 days post-GBCA administration was performed here based on the same LA-ICP-MS measurements. The same applies to the *in vivo* $T_1$ relaxometry data acquired 10 days post-GBCA administration (see Figure SDC_4, Supplemental Digital Content 3 for further details). The setup of the *in vivo* trial included EAE and HC mice of the same age, sex, and strain, and followed the same GBCA administration regimen as applied in (47).

Third, we compared the calibrated $T_1$ values to these *in vivo* $T_1$ ($T_{1,iv}$) and LA-ICP-MS data. For this, we calculated values $R_{1,c}(C_{Gd}) = r_1 \times C_{Gd} + R_1(0)$, i.e. expected relaxation rates at the LA-ICP-MS-detected Gd concentration within the different brain regions. Thereby, $R_1(0)$ is the averaged inverse of the $T_1$ relaxation times measured within each brain region in EAE and HC mice before the first GBCA administration (GBCA baseline, corresponding to day 12 of the experimental setup shown in Figure 1, respectively. The percentage $T_1$ deviation of *in vivo* $T_{1,iv}$ from calculated $T_{1,c}$ (=$1/R_{1,c}$) is then ($T_{1,c}-T_{1,iv})/T_{1,c} \times 100$. Negative $T_1$ deviation corresponds to longer $T_1$ relaxation times than expected for the Gd concentration. This can be expressed either by a reduced effective relaxivity $r_{1,eff} = (R_{1,c} - R_1(0))/C_{Gd}$ at the LA-ICP-MS Gd concentration or by a reduced effective Gd concentration $C_{Gd,eff} = (1/T_{1,iv} - R_1(0))/r_1$ with the *in vitro* calibrated $r_1$. The respective $R_1(0)$ and $T_{1,iv}$ values are detailed in the Supplemental Digital Content 3.

## 2.11 LA-ICP-MS

For the LA-ICP-MS measurements an NWR-213 laser ablation system equipped with a 2-volume sample chamber (Elemental Scientific Lasers, Bozeman, MT, USA) was coupled to an ICP sector-field mass spectrometer (Element XR; Thermo Fisher Scientific, Bremen, Germany). Cerebellar cryosections (thickness 10 µm) of EAE and HC mice mounted on microscopic glass slide (SuperFrost Plus adhesion slides, Thermo Fisher Scientific, Schwerte, Germany) were ablated in an imaging mode. For the Gd quantification, agarose gels spiked with different Gd concentrations were used. Further details are given in Anderhalten *et al.* (47). Data visualization was done in Origin 2018 (OriginLab Corporation, Northampton, MA, USA). The measured values for the isotopes $^{158}$Gd and $^{55}$Mn were used for the evaluation. Since the samples and the calibration standards based on spiked agarose have a natural isotope abundance, the Gd and Mn images reflect the quantitative distribution of Gd and Mn, respectively, (in µM) in the thin tissue sections.

To determine the Gd retention in certain brain areas (medulla, cerebellar cortex and CN) manual ROIs were applied to the calibrated LA-ICP-MS images for Gd. The analysis of each ROI was conducted 3 times in an independent randomized manner using ImageJ (National Institute of Health, USA) for better reliability. Gd concentrations determined for the medulla, cerebellar cortex and CN of EAE and HC mice 10 days after GBCA injection were averaged.

## 2.12 Statistics

The study data were statistically analyzed using R software (RStudio, version 2022.07.2). LA-ICP-MS and MRI calibration data were expressed descriptively (*n*=1-2/group or condition, each) as geometric means ± standard deviations (SD). Concentration results obtained by cwEPR were given descriptively as total numbers, and results derived from $^{31}$P ENDOR fits were stated as value ± SD or ± a resolution limit if the uncertainty provided by the fit was unreasonable (see Supplemental Digital Content 2). *In vivo* MRI data for the three brain biopsy regions (*n*=8-10/group) were analyzed by applying non-parametric Mann-Whitney U tests for two-group comparison, followed by Bonferroni correction for multiple testing (see Supplemental Digital Content 3). P values < 0.05 indicated statistical significance with * implying p < 0.05, ** implying p < 0.01, and *** implying p < 0.001.

## 3. RESULTS

### 3.1 Gd retention in brain tissue after *in vivo* and *ex vivo* administration of GBCA

To assess the presence of Gd retention and to evaluate the potential release of $Gd^{3+}$ from the GBCA complex, we recorded continuous wave (cw)EPR spectra of pure standards, the biopsies obtained 10 days after multiple i.v. injections of GBCAs (Figure 3a, b) and the slices following 48 h-GBCA treatment *ex vivo* (Figure 3c, d). Figures 3a and 3c depict the experimental spectra of GBCA-treated samples (black) in comparison to the respective negative control samples (blue). The spectra are dominated by six lines, which are due to traces of $Mn^{2+}$ typically present in cell and tissue samples, and was also detected by LA-ICP-MS in the tissue thin sections of the mouse brains (47). A broader (in part barely visible) feature on the high field side of the fifth $Mn^{2+}$ line (from the left) represents the $Gd^{3+}$ signal. Figures 3b and 3d contain difference spectra between the GBCA-treated biopsy or slice samples of Figures 3a and 3c and their negative controls. The subtraction removed most of the $Mn^{2+}$ signal, but left residuals around the $Mn^{2+}$ line positions (marked with asterisks). The difference spectra (red) are displayed together with correspondingly scaled pure GBCA reference spectra (yellow).

We detected similar amounts of retained Gd in the range of 2–3 µM for both GBCAs in the biopsies from different EAE cerebellar regions (cerebellar cortex, CN, and medulla). The spectra of biopsies from the mouse treated with the linear GBCA gadopentetate EAE mouse differ visually from the pure GBCA spectrum (Figure 3b). Here, a distinct narrow line contribution is clearly discernible in the difference spectrum, particularly visible in the negative spectral part. Moreover, we were able to decompose the spectrum of the retained Gd in the biopsies into a spectrum arising from intact gadopentetate and a spectrum of free $Gd^{3+}$ with a ratio of 7:3, as shown in Figure SDC_2 (see Supplemental Digital Content 1).

The cortical biopsy of the EAE mouse treated with the macrocyclic GBCA gadobutrol shows a spectrum similar to the pure gadobutrol spectrum, which is not distinct enough from the free $Gd^{3+}$ spectrum to be used for deconvolution (Figure 3b, also see Supplemental Digital Content 4, showing gadobutrol-derived cwEPR spectra in Figure SDC_5 versus $^1$H ENDOR spectra in Figure SDC_6). It should be noted that the larger signal amplitude of the EAE cortex biopsy post-gadobutrol compared to post-gadopentetate is misleading in terms of Gd concentration due to the significantly lower line width and both samples contain similar Gd concentrations. The cortical biopsy of the gadobutrol-treated HC mouse did not show any Gd signal (Figure 3b).

The Gd spectra obtained from slices incubated with GBCAs *ex vivo* (Figure 3c) are within the noise range, identical to the spectrum of the respective pure GBCA (Figure 3d). Their intensities are up to 5-fold the signals of the Gd in the biopsies, corresponding to ~8-10 µM.

## 3.2 Environment of retained Gd after *in vivo* GBCA administration

To further elucidate whether the retained Gd found in the different samples after GBCA administration was still bound within the parent GBCA complex, as well as to gain deeper insights into retention mechanisms and potential competing Gd binding sites within the tissues, we directly measured the local chemical environment of Gd using electron-nuclear double resonance (ENDOR). The $^1$H ENDOR spectra of the cortical biopsies and slice samples are shown in Figure 4 (in black) in comparison to the spectra of the pure GBCAs (in yellow). The spectrum shown in blue in Figure 4a was recorded on the $Mn^{2+}$ EPR signal part and represents $Mn^{2+}$-H hyperfine couplings. The absence of any significant intensity around ±3.5 MHz in the spectrum of the gadopentetate biopsy sample (Figure 4a, black) recorded on the $Gd^{3+}$ EPR signal shows the achieved efficient $Mn^{2+}$ suppression. Thus, all spectra depicted in black correspond exclusively to $Gd^{3+}$-H hyperfine couplings.

The spectrum of the Gd couplings of the sample from the gadopentetate-treated EAE mouse clearly differs from the one of pure gadopentetate. In contrast, the spectra recorded for the gadobutrol-treated EAE mouse (Figure 4b) as well as for the *ex vivo* slices treated with both gadopentetate (Figure 4c) and gadobutrol (Figure 4d) coincide with the spectra of the corresponding pure GBCA. The small differences in the center of the obtained spectra (below ±0.4 MHz)), which correspond to very weak couplings and, therefore, large distances (Figure 4b,c,d), can be neglected here. The pronounced spectral difference in the case of the gadopentetate-treated EAE mouse (Figure 4a) demonstrates that at least a fraction of the retained Gd exists in an environment different from the GBCA complex cage.

3.3 Phosphorus in the vicinity of retained Gd

Since phosphorus occurs with 100 % natural abundance as the $^{31}$P magnetic isotope, it is an interesting species to probe in the vicinity of the retained Gd by ENDOR, even though Gd in the inorganic, solid $GdPO_4$ is invisible to EPR methods. Multiple organic P-containing species, e.g. organic phosphates in nucleotides, phospholipids or proteins/sugars with phosphorylated residues are potential binding sites. The P ENDOR spectra of the cortical biopsy from the gadopentetate-treated EAE mouse and of the *ex vivo* slices treated with both GBCA are shown in Figure 5. The spectrum of the cortical biopsy (black, first plot) reveals distinct peaks at about ±0.4 MHz indicative for P in the first coordination shell of Gd. P in the vicinity of $Gd^{3+}$ was also detected in the GBCA-treated *ex vivo* slices for both GBCA. Additional $^{31}$P Mims ENDOR were recorded for the reference samples Gd-ATP (green), Gd-DMPC (magenta), and gadopentetate-ATP (black). The spectra of the control samples of Gd-ATP (green) and gadopentetate-ATP (black) shown at the bottom are indistinguishable except for the much worse signal-to-noise ratio (SNR) for the gadopentetate-ATP sample at equal Gd concentrations.

The splitting between the peaks shown in Figure 5 encodes the hyperfine coupling between the Gd electron spin and the $^{31}$P nuclear spin and provides a measure for the Gd-P distance. This splitting appears slightly larger for the EAE biopsy compared to the Gd-ATP reference sample (green) and seems to be in better agreement with the Gd-DMPC reference sample (magenta) which shows the largest peak splitting of the reference samples, i.e. the shortest Gd-P distance. The slightly larger peak splitting for the EAE biopsy indicates a slightly shorter Gd-P distance than in Gd-ATP. The peak splittings are smaller for the slice samples than for Gd-ATP, indicating a larger Gd-P distance in these cases.

We extracted the $^{31}$P hyperfine couplings and the respective Gd-P distances $r_{Gd-P}$ by least-square fitting (see Table SDC_1, Supplemental Digital Content 2 for full P ENDOR fit results). The Gd-P distance obtained for the Gd-ATP reference is $r_{Gd-P}$ = 0.36 ± 0.01 nm, for the gadopentetate-treated EAE mouse sample $r_{Gd-P}$ = 0.35 ± 0.01 nm, for the *ex vivo* slice samples $r_{Gd-P}$ = 0.37 ± 0.01 nm, and for the Gd-DMPC reference $r_{Gd-P}$ = 0.36 ± 0.03 nm due to the low SNR. The $r_{Gd-P}$ determined for Gd-ATP here is slightly larger than the $r_{Gd-P}$ = 0.35 obtained from NMR relaxation (34), however, it agrees with recent ENDOR data (65), and in (34) also the Mn-P distance for Mn-ATP was significantly shorter than the distance deduced from ENDOR (66). A P ENDOR signal was measurable neither for the cortical biopsy sample of the gadobutrol-treated EAE mouse nor for a gadobutrol-ATP reference sample.

3.4 Interaction of Gd with inorganic ligands

We performed an *in vitro* MRI calibration experiment to reveal the relaxation active amount of bound and unbound Gd retained in the tissue. The relaxivity $r_1$ of gadopentetate and gadobutrol was computed using *in vitro* $T_1$ relaxometry, with standards containing ascending Gd$^{3+}$ concentrations in water, FBS, and homogenized brain tissue (see Table SDC_2, Supplemental Digital Content 3 for detailed results). Tested standards demonstrated a good fit to the linear regression model with relaxivities $r_1$(gadopentetate) = 4.35 s$^{-1}$mM$^{-1}$ and $r_1$(gadobutrol) = 6.96 s$^{-1}$mM$^{-1}$ (homogenized brain; gadopentetate, $R^2$ = 0.97; gadobutrol, $R^2$ = 0.98).

The computation of $T_1$ deviation (%) comparing *in vivo* $T_1$ values in the brain to calibrated $T_1$ values in brain homogenate at the respective Gd concentrations detected by LA-ICP-MS was performed for the three biopsy regions medulla, cerebellar cortex, and CN of EAE and HC mice (Figure 6). The *in vivo* $T_1$ relaxometry and LA-ICP-MS data underlying our following results are reported here as part of the Supplemental Digital Content 3 (Figure SDC_4g-i and Table SDC_3), with only the CN Gd concentrations having been previously published in (47). For better traceability of the $T_1$ deviation computation, here, the CN data at day 10 post-GBCA application are shown again, as illustrated in Figure SDC_4. Mean $T_1$ relaxation times measured *in vivo* 10 days post-GBCA *in vivo* administration deviate only slightly from calibrated values inside the brain regions medulla (Figure 6a) and the cerebellar cortex (Figure 6b) for both EAE and HC mice and both GBCA. Inside the CN, the same holds true for both EAE and HC mice but only after gadobutrol treatment (Figure 6c). In contrast, a strong negative $T_1$ deviation (Figure 6c) was observed after gadopentetate treatment, i.e. the $T_1$ relaxation times measured *in vivo* post-GBCA administration were longer than the corresponding *in vitro* calibrated $T_1$ (see Supplemental Digital Content 3 for further explanation). Thus, the retained Gd inside the CN 10 days after repeated gadopentetate administration lowers $T_1$ in *in vivo* MRI less than expected for the LA-ICP-MS determined concentration. This effect was about 1.5-fold stronger in EAE compared to HC mice (mean ± SD; EAE vs HC; CN: –22.4 ± 8.3 % vs –14.8 ± 3.1 %). The relaxation time deviation is consistent with either a markedly reduced relaxivity $r_{1,\text{eff}}$ (EAE vs HC; CN: 1.45 s$^{-1}$mM$^{-1}$ vs 0.78 s$^{-1}$mM$^{-1}$) compared to the *in vitro* result in homogenized brain tissue ($r_{1,\text{eff}}$ = 4.35 s$^{-1}$mM$^{-1}$), or with a severely reduced effective average Gd concentration $C_{\text{Gd,eff}}$ (EAE vs HC; CN: 14.4 µM vs 3.7 µM) compared to the absolute LA-ICP-MS-detected Gd concentrations $C_{\text{Gd}}$ ((47): mean ± SD; EAE vs HC; CN: 42.60 ± 10.57 µM vs 21.28 ± 3.72 µM; also see Table SDC_3, Supplemental Digital Content 3 for full results).

## 4. DISCUSSION

We show that *in vivo* Gd retention after GBCA treatment can be detected by cwEPR and pulsed ENDOR on cerebellar cortex biopsy samples as well as chronic organotypic hippocampal tissue slices, both of sub-mm dimensions. Tissue biopsies were collected from different regions of the cerebellum of EAE and HC mice 10 days after the final i.v. administration of a linear or a macrocyclic GBCA. In all EAE biopsies, the Gd retention range was about 2–3 µM as determined by cwEPR, regardless of the GBCA used. However, Gd³⁺ release was detected only from the linear gadopentetate in biopsy samples but not in organotypic brain slices, where both gadopentetate and gadobutrol remained intact. ³¹P ENDOR revealed P near the released Gd³⁺ after *in vivo* administration of linear gadopentetate. These results are relevant in several respects. First, they show the capability of EPR to detect Gd retention on a biopsy level and second, the EPR data corroborate that neuroinflammation promotes Gd retention in the CNS after multiple injections of GBCA, as demonstrated in a previous study (47).

The lower concentration limit of 2 µM retained Gd estimated from cwEPR in the gadobutrol-treated EAE cerebellar cortex biopsy is in line with the concentrations detected previously by LA-ICP-MS (mean value cerebellar cortex, granular layers only, 1.6 µM) (47), and specifically for the cerebellar cortex from the same specimen as the EPR sample with 2 µM Gd. Ten days after the final administration of gadopentetate, the mean Gd content of EAE mice was 43 µM in the CN (47), 13 µM in the cerebellar cortex, and 10 µM in the medulla, as quantified here from previously acquired LA-ICP-MS data (Figure SDC_4g-I, Supplemental Digital Content 3). The specific specimen of the EPR samples reached even 50 µM in the CN, 17 µM in the cerebellar cortex, and 13 µM in the medulla, which is significantly more than the retention levels estimated by cwEPR (CN 3 µM, cerebellar cortex 3 µM, and medulla 2 µM). With this, two clear discrepancies between the EPR and LA-ICP-MS data are visible. First, the concentration range determined by LA-ICP-MS is significantly larger than that from EPR and second, LA-ICP-MS shows a strong difference between CN (47) and cerebellar cortex as well as medulla, while EPR detects similar concentrations. The latter goes in parallel with the findings from the calibrated *in vivo* MRI. There, we also found that the $T_1$ shortening effect within the CN is strongly reduced with respect to the retained Gd amount in HC and EAE mice, while it roughly corresponds to the concentrations detected by LA-ICP-MS in the cerebellar cortex and the medulla. This aspect will be discussed in more detail below.

Other than the absolute Gd concentration, further important information can be taken from the analysis of the cwEPR spectra. After the *in vivo* gadopentetate treatment of the EAE mouse, two distinct Gd species can be separated. One component is identical to the pure

GBCA spectrum and the second component is well represented by the spectrum of $Gd^{3+}$ in aqueous solution. Quantitative analysis of the spectra revealed that roughly 70% correspond to intact GBCA complexes and the rest to $Gd^{3+}$ in a different environment, resulting in magnetic parameters similar to those in aqueous solution (see Figure SDC_2, Supplemental Digital Content 1). The occurrence of $Gd^{3+}$ outside the GBCA complex supports the assumption that at least part of the administered Gd was released *in vivo* from the chelate after repeated gadopentetate injections. $Gd^{3+}$ release is more likely from the kinetically less stable linear gadopentetate (22, 29, 30), in line with the much larger Gd concentrations found by LA-ICP-MS for gadopentetate than for the macrocyclic gadobutrol (47). However, cwEPR is inconclusive with respect to a possible $Gd^{3+}$ release from gadobutrol since the pure gadobutrol spectrum and the $GdCl_3$ ($Gd^{3+}$) reference spectrum are too similar for a distinction (see Figure SDC_5, Supplemental Digital Content 4).

The *ex vivo* study on chronic organotypic hippocampal slices aimed to evaluate Gd binding stability over time within the tissue. Slices were pretreated with the inflammatory mediator TNFα, as our previous data indicated that an inflammatory milieu may promote gadopentetate-related cytotoxicity by presumably promoting $Gd^{3+}$ release (47). After the GBCA treatment, the slices were cultured for one more week under native conditions, and subsequently washed and homogenized to remove non-bound Gd species. Gd could be detected in tissue homogenate of GBCA-treated slices after repeated steps of homogenization and washing, suggesting a high binding stability of the detected Gd species. The concentration of Gd retained after incubation with both GBCAs lies between 8 and 10 µM, corresponding to about 0.8–1 % of the originally administered GBCA dose. Interestingly, the spectra for both utilized GBCAs were highly similar to the pure GBCA spectra, thus, no evidence for $Gd^{3+}$ release was found in the slices, in contrast to the biopsy sample, where release from gadopentetate was concluded from cwEPR due to clear spectral differences.

Moreover, we utilized $^1H$ ENDOR for further investigation of $Gd^{3+}$ release from the linear gadopentetate, as suggested by the cwEPR results on the biopsy samples, and to assess a possible release from the macrocyclic gadobutrol that was indistinguishable by cwEPR. These spectra provide information on protons in the vicinity of Gd in terms of hyperfine coupling, which is connected to Gd–H distances. The difference observed between the spectra of the biopsy of the gadopentetate-treated EAE mouse and the one of pure gadopentetate demonstrates a clear variance in the H environment of Gd and supports the finding of partial $Gd^{3+}$ release in biopsies from the gadopentetate-treated EAE mouse, as concluded from the cwEPR data. In contrast, the spectrum of the gadopentetate-treated slices is well comparable to the spectrum of pure gadopentetate and no release of $Gd^{3+}$ from gadopentetate can be seen in the slices, but retention of the intact GBCA complex.

The spectra of the biopsy of the EAE mouse treated with gadobutrol as well as of those of the chronic slices are all highly similar to those of the pure corresponding GBCAs with only minor intensity changes for distant H atoms, indicating an intact GBCA structure. Furthermore, the $^1$H ENDOR spectra of $Gd^{3+}$ in aqueous solution and in gadobutrol are clearly distinct (see Figure SDC_6, Supplemental Digital Content 4) while their cwEPR spectra are virtually indistinguishable. A discernable contribution of the ENDOR spectrum for free $Gd^{3+}$ is present neither in the biopsy sample of the gadobutrol-treated EAE mouse nor in the chronic slices. This further supports that Gd retention detected in both tissue types after gadobutrol treatment occurs in the intact GBCA structure. The striking difference between the $^1$H ENDOR spectra of the biopsies after gadopentetate and gadobutrol administration clearly evidences the power of the method for detecting $Gd^{3+}$ release from GBCA in tissue.

Protons are very abundant and therefore the change in H environment does not provide sufficient information about new Gd binding sites. It is known that $Gd^{3+}$ has a high affinity to phosphates with a solubility product of around $10^{-25}$ (67) and phosphorus close to Gd accumulation was observed before (24, 68, 69). In autopsy skin tissues from a patient with nephrogenic systemic fibrosis treated with GBCA (69), P at a distance of 0.31 nm and 0.37 nm from Gd and a Gd-Gd distance of 0.4 nm using X-ray fluorescence (SXRF) microscopy and extended X-ray absorption fine structure (EXAFS) spectroscopy was determined. However, the authors have attributed their findings to a composition of Na, Ca, Gd and P, consistent with insoluble $GdPO_4$ (69), which is not visible in EPR.

Gd-P interactions are well detectable with ENDOR due to the magnetic moment of $^{31}$P. Using P ENDOR, we demonstrated in this study that there is indeed P in the Gd vicinity. The absence of these P signals in spectra taken on the EPR signal of the unavoidable $Mn^{2+}$ contamination in the gadopentetate-treated EAE mouse additionally confirms the efficacy of the spectral separation between $Gd^{3+}$ and $Mn^{2+}$ (see Figure SDC_7, Supplemental Digital Content 5, detailing $Mn^{2+}$ suppression on P ENDOR spectra). The line shape for the Gd-P couplings measured on the biopsy of the gadopentetate-treated EAE mouse appears to be different and the peak splitting is larger than in the Gd-ATP reference sample. The splitting corresponds to a Gd-P distance of 0.35 ± 0.01 nm, slightly smaller than the 0.36 nm for the reference sample Gd-ATP. Therefore, another reference measurement with DMPC, which belongs to one of the most abundant lipid classes in mouse brain tissue (70) and contains only one phosphate group, was performed. Lanthanides have been shown to bind to phospholipid membranes before (71, 72). The P ENDOR data suggest a somewhat larger splitting than for Gd-ATP, the distances obtained from spectral fits, however, coincide within the error range. The SNR of the Gd-DMPC is much worse than anticipated for the $Gd^{3+}$ concentration when compared to the Gd-ATP reference, while the EPR signal intensity was as expected, and indicates a much weaker Gd-DMPC complexation. A similar effect was found for the binding of adenine monophosphate (AMP) to $Gd^{3+}$, which was markedly weaker compared to adenosine diphosphate (ADP) and ATP (65). This can be understood by the entropic effect of the bidentate binding of ATP vs. the monodentate binding motif of AMP and DMPC to $Gd^{3+}$. The high binding affinity of ATP to $Gd^{3+}$ has been shown to compete with the Gd-DTPA complex of gadopentetate for the paramagnetic ion and direct Gd complexation by ATP occurs in ATP/gadopentetate mixtures, while a ternary DTPA-ATP complex has been excluded (73). This fits well the virtually identical P ENDOR spectra of the Gd-ATP and the gadopentetate-ATP sample, and the strongly reduced SNR in the gadopentetate-ATP sample reflects the strong prevalence of Gd-DTPA over Gd-ATP.

The splitting in the P ENDOR measurements of the chronic brain slices incubated with both GBCA is visibly smaller and corresponds to a Gd-P distance of 0.37 ± 0.01 nm. From cwEPR and $^1$H ENDOR measurements, we concluded that both GBCA retained in slices stayed intact. Therefore, it was surprising to see P atoms in such a vicinity to Gd. A possible explanation is that a phosphate group might occupy the water exchange site of the GBCA. ATP can be excluded as a binding partner since a gadobutrol-ATP reference sample yielded no detectable P ENDOR spectrum. Possible candidates would again be phospholipids (71, 72) as well as proteins or sugars with phosphorylated residues, including components of the ECM (74, 75).

The biopsy sample of the gadobutrol-treated EAE mouse, which, analogously to the incubated slices, contained only intact GBCA as determined by $^1$H ENDOR, showed no P in the close vicinity of Gd, just as the gadobutrol-ATP reference sample. This observation suggests that GBCA applied directly to the surface of tissue slices may exhibit different behavior compared to GBCA administered intravenously *in vivo* and subsequently detected in biopsies. *In vivo*, the contrast agents reach the tissue only after interacting with physiological transport and barrier structures and molecules, such as serum proteins (32, 35) or vasculature-associated cellular and ECM components (27, 33, 35, 36, 76).

*In vitro* MRI calibration revealed that the $T_1$ relaxation time observed by *in vivo* MRI within the CN 10 days post-gadopentetate administration in HC and, particularly, in EAE mice was significantly longer than in homogenized brain tissue spiked with the Gd concentration as detected by LA-ICP-MS. In contrast, *in vivo* MRI showed the $T_1$ shortening effect expected at the Gd concentration determined by LA-ICP-MS 10 days post-gadobutrol administration in both EAE and HC brains. The observed $T_1$ deviation post-gadopentetate administration indicates that the relaxation-active Gd constitutes only a fraction of the total Gd amount present, in line with the findings that the largest portion of retained Gd in rat brain occurs as deposits, mainly composed of GdPO$_4$ (24, 32). Such inorganic GdPO$_4$ salt deposits likely arise due to their low solubility, which is a possible reason for a weak relaxivity in MRI and loss of EPR signal. It should be mentioned that LA-ICP-MS detects the entire Gd present in the tissue section; a differentiation of species or binding forms is not possible with this method. Thus, our data suggests that the presence of Gd deposits after gadopentetate administration may cause both the $T_1$ deviation in MRI and the drastically reduced EPR-visible Gd amount. Formation of the inorganic Gd deposits requires a release of the $Gd^{3+}$ from the GBCA chelate, which is more likely from the kinetically less stable linear gadopentetate (47). No substantial $Gd^{3+}$ release and thereby no inorganic deposits seem to occur for gadobutrol.

However, a significant discrepancy still exists between the Gd concentrations post-gadopentetate when comparing the healthy to the EAE mouse. During EAE, the formation of insoluble precipitates after i.v. gadopentetate injection may be additionally promoted by inflammation-related changes in endogenous ion homeostasis (e.g., zinc) and reduced chelate stability (47, 77). Moreover, in line with our findings, after linear GBCA administration in humans, insoluble deposits were particularly found in CN specimens, within capillary walls (29, 50) or neuronal nuclei (48). The CN contain a high density of neuronal bodies that are metabolically active, potentially leading to accumulation of inorganic phosphates. This, however, is speculative, and reasons why the CN region is a predilection site for the formation of deposits with low relaxivity remain to be elucidated.

While such inorganic deposits are invisible to EPR, it does detect retained Gd that either remained in intact GBCA complex or is bound to alternative complexing species. The distinct shapes of cwEPR spectra for the biopsies of the gadopentetate-treated EAE mouse and pure gadopentetate support the assumption that at least part of the detectable Gd was released from the chelate after repeated *in vivo* injections and complexed by organic ligands. Quantitative analysis of the spectra revealed that roughly 70% of the EPR-visible Gd is still present as intact GBCA while the other 30% is found in a different, likely macromolecular environment (see Figure SDC_2, Supplemental Digital Content 1 for visualization). Binding of $Gd^{3+}$ to macromolecular moieties might be a reason for differences in the relaxation-effective Gd concentration as determined by calibrated MRI and the EPR-visible Gd concentration. The relaxivity of $Gd^{3+}$ bound to macromolecules can be significantly higher than for GBCA complexes (78, 79). A relaxivity enhancement by a factor of 5 was previously observed by others (80). Such increased relaxivity of the 30 % fraction of Gd found outside the GBCA would lead to an overestimation of the relaxation-active Gd concentration by more than a factor of two and thereby significantly underestimate the amount of Gd in inorganic deposits. However, this would still only explain part of the of Gd missing in EPR and requires further exploration.

## 5. CONCLUSIONS

In summary, we have shown EPR to be a powerful tool for investigating paramagnetic species and their local environment inside intact tissue structures. Importantly, we successfully separated the Gd signal from the significantly stronger $Mn^{2+}$ signal, a common challenge in EPR measurements of biological samples. CwEPR revealed Gd accumulations in EAE cerebella *in vivo* and brain slice cultures *ex vivo* after administration of both GBCAs, linear gadopentetate and macrocyclic gadobutrol. In contrast, no Gd was detectable in the gadobutrol-treated HC cerebellum, confirming previous findings on inflammation-promoted Gd retention. Moreover, $Gd^{3+}$ release was exclusively detected after *in vivo* linear GBCA administration, but not within organotypic brain slices.

The proof-of-concept experimental design, including frequency and daily GBCA doses, differed significantly from applications in clinical practice. However, for consistency, the chosen treatment regimen adhered to our previous publications (46, 47). The low sample size of one mouse per condition for the *in vivo* trial, as well as one final tissue homogenate per *ex vivo* condition, precluded quantitative statistical analysis, limiting the ability to draw statistically valid conclusions regarding $Gd^{3+}$ retention and release during neuroinflammation.

Furthermore, organotypic brain slice cultures do not always permit direct comparisons with experiments *in vivo*, in which variables such as the blood-brain barrier, blood factors that could influence compound stability, and directed elimination of GBCAs (32) may additionally influence the kinetics of Gd retention. *Ex vivo*, a successful chronic slice culture was not feasible without the use of P-containing buffer (PBS), which in turn could affect our findings on $Gd^{3+}$ release and binding.

ENDOR measurements revealed a chelation of the released $Gd^{3+}$ in the cerebellar biopsy by phosphorus-containing moieties. Phosphorus was also found in the surrounding of the intact gadopentetate and gadobutrol in the tissue slices. These findings underscore the complexity of GBCA interactions in brain tissue and highlight the importance of integrating *in vivo* and *ex* *vivo* approaches differences between *in vivo* and *ex vivo* models to elucidate the mechanisms underlying sustained Gd retention. Thereby, the detection of retained Gd and the distinction between intact GBCA and released species on the biopsy level by EPR/ENDOR opens a new route for organ and tissue-specific studies.

Lastly, our MRI *in vitro* calibration experiments suggest that a substantial portion of retained Gd within the CN may be undetected by *in vivo* MRI after repeated gadopentetate administration, particularly in the context of neuroinflammation. While MRI remains one of the best diagnostic tools for brain pathologies in clinical practice, the potential underestimation of the true extent of long-lasting Gd retention in specific brain regions is of considerable clinical importance, particularly for patients with neuroinflammatory conditions.

**Figures:**

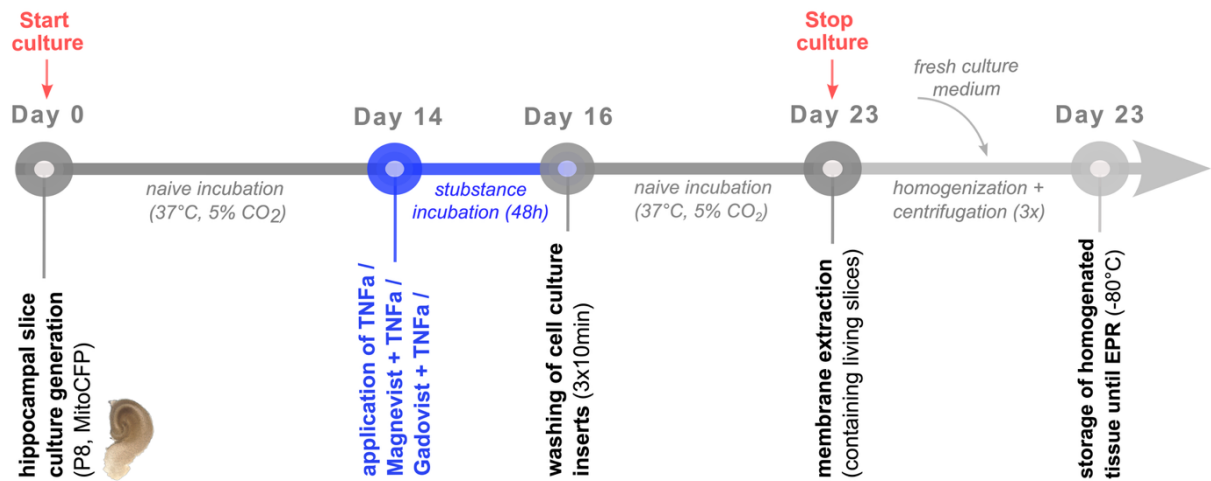

**Figure 1. a)** Experimental mouse setup for EPR sampling with repeated *in vivo* injections of either gadopentetate (*n*=1) or gadobutrol (*n*=1) in EAE mice as well as gadobutrol in HC (*n*=1) at 2.5 mmol/kg BW each (eight daily injections from day 13 to 23 with a two-day pause in between; cumulative dose of 20 mmol/kg BW). The GBCA-treated mice and one untreated HC mouse as negative control were sacrificed 10 days post-GBCA administration and brains were processed for EPR measurements. This simplified experimental setup is based on the setup previously published in (47). **b)** Exemplary visualization of the location of brain biopsies for EPR measurements in a cerebellar hematoxylin and eosin-stained EAE mouse brain section at 2x magnification. Blue outlines indicate the position of biopsies from the cerebellar cortex, red outlines indicate CN, and black outlines indicate the medulla. (Scale bar 1 mm).

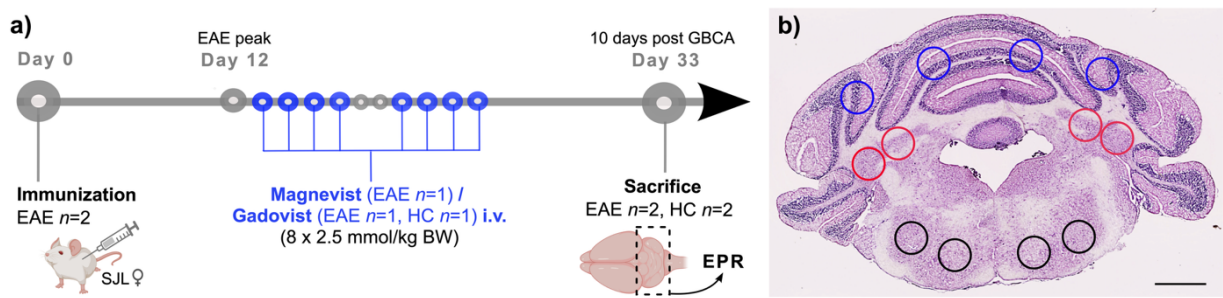

**Figure 2.** Illustration of the experimental *ex vivo* setup. Chronic organotypic hippocampal cultures were exposed for 48 h to GBCAs and TNF starting on day 14 post-preparation (3 conditions: *1.* TNFα at 50 ng/ml, 2. 1 mM gadopentetate + TNFα, *3.* 1 mM gadobutrol + TNFα). On day 16, slices were washed several times and cultured for an additional week without substances. On day 23, culture membranes containing hippocampal slices were carefully extracted. After repeated homogenization and centrifugation steps with fresh culture medium (3 times, 0.5 ml medium each), one final tissue homogenate per condition was obtained.

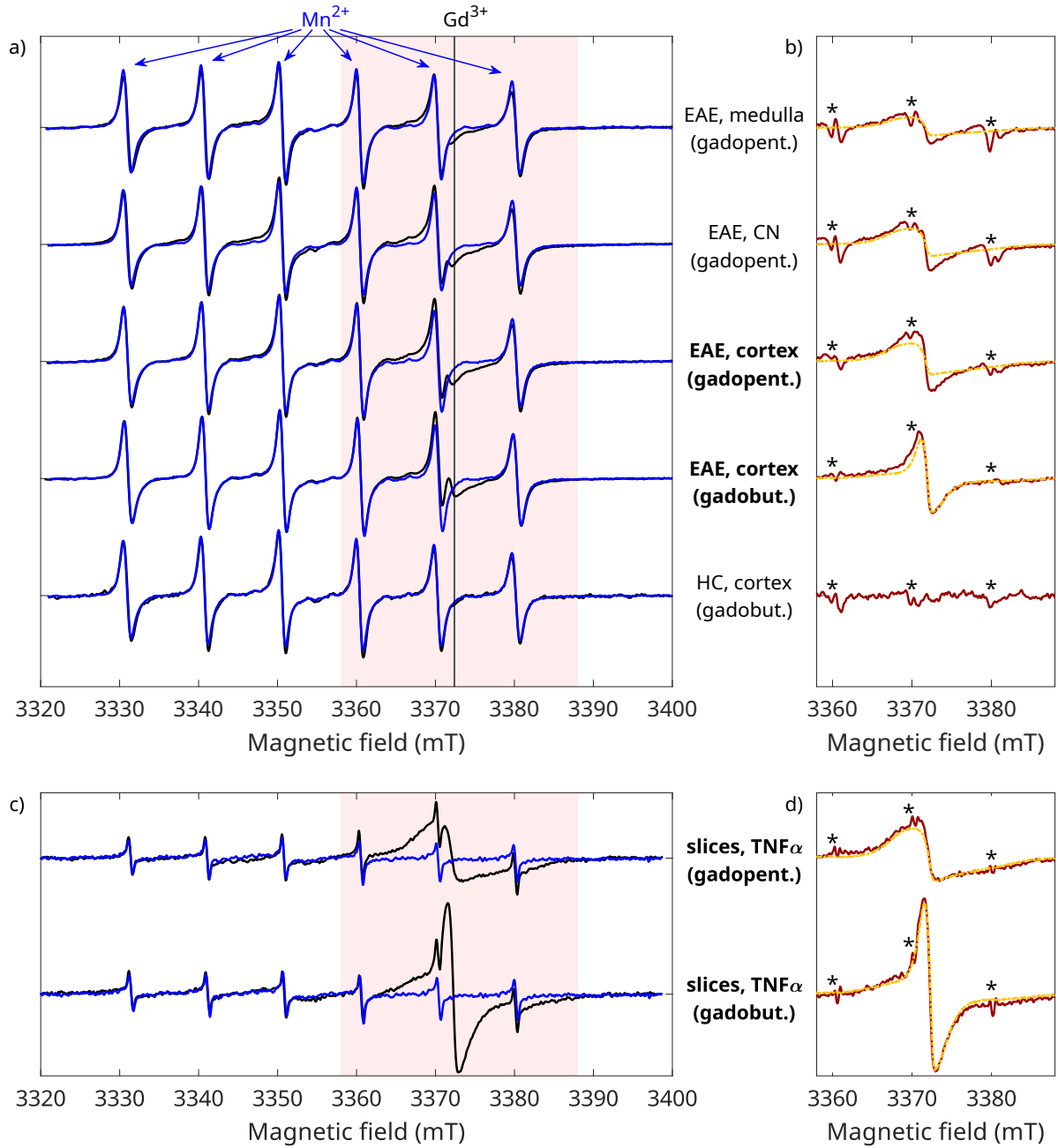

**Figure 3.** Comparison of cwEPR spectra. **a)** Experimental cwEPR spectra (in black) of biopsies from different brain regions (medulla, CN, and/or exclusively cerebellar cortex) obtained from EAE mice treated with either gadopentetate (linear GBCA) or gadobutrol (macrocyclic), alongside one from a HC mouse treated with gadobutrol are compared to normalized control sample spectra (in blue). The area highlighted in red corresponds to the field range evaluated in **b)**. There, differences between the blue and black spectra (all scaled by a factor of 2 compared to **a**) are plotted in red and compared to the scaled corresponding pure GBCA spectra (yellow). Residual $Mn^{2+}$ artifacts are marked with asterisks. **c)** Experimental cwEPR spectra of chronic slices incubated with gadopentetate or gadobutrol *ex vivo*, respectively, are compared to a negative control spectrum, along with **d)** corresponding difference spectra compared to pure GBCA spectra. The four samples indicated in bold font were additionally measured with ENDOR.

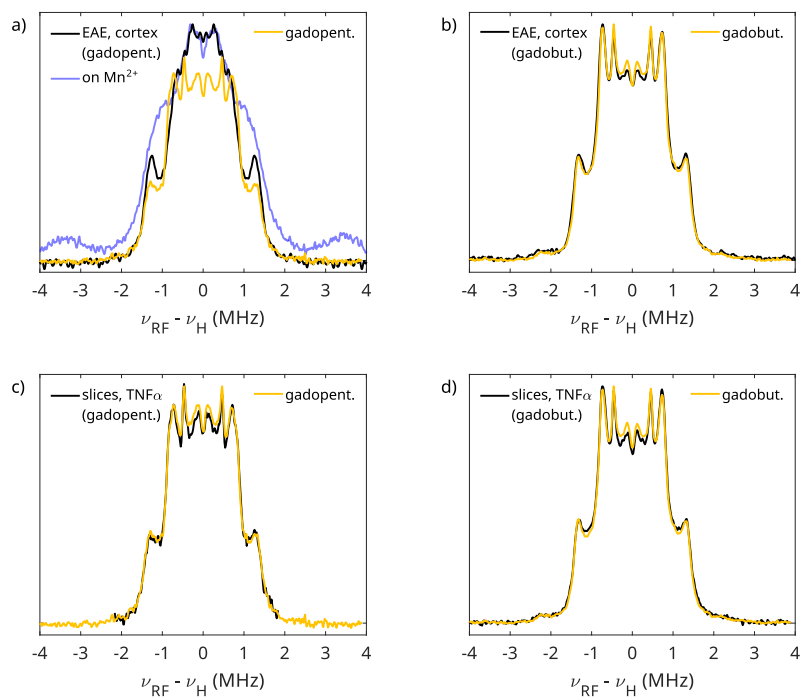

**Figure 4**. $^1$H ENDOR spectra recorded on **a), b)** cortical mouse brain biopsies (black) and **c), d)** slices incubated with GBCA *ex vivo* (black), compared to the normalized spectra of the corresponding pure GBCA (shown in yellow). In **a)**, the spectrum for the $Mn^{2+}$-H couplings is shown in blue, recorded on the lowest-field $Mn^{2+}$ line, where the Gd signal is negligible. In **a)**, the spectrum of pure gadopentetate was scaled for comparison; all other spectra were normalized to their maximum.

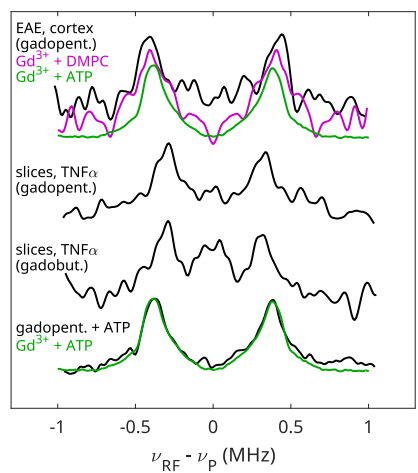

**Figure 5**. P ENDOR spectra recorded on the cortical mouse brain biopsy of the gadopentetate-treated EAE mouse and the slices incubated *ex vivo* (black; three upper plots). As references, the spectra of Gd-ATP (green), Gd-DMPC (magenta, symmetrized data) and gadopentetate-ATP are shown. Gd-DMPC and Gd-ATP are slightly shifted downwards compared to the biopsy spectrum (top) for clarity. All spectra are normalized to the maximum of the smoothed data. The spectra of the gadopentetate-treated EAE mouse and slices were measured with Mn suppression, the gadobutrol-treated slices as well as all the reference samples were measured with the standard sequence.

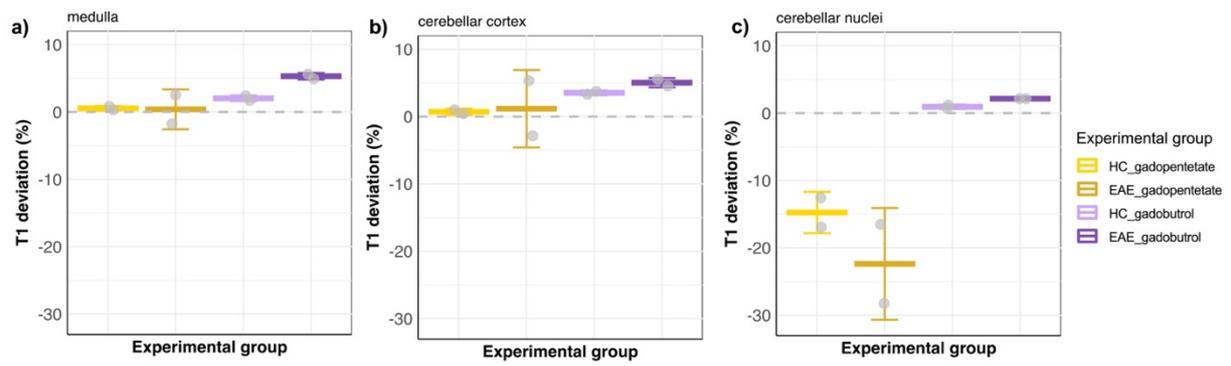

**Figure 6.** $T_1$ deviation (%) 10 days post-GBCA administration is displayed for **a)** the medulla, **b)** the cerebellar cortex, and **c)** the CN for both tested GBCAs. The crossbars represent the mean, and whiskers indicate ± 1 SD. Within the CN area, *in vivo* $T_1$ values after gadopentetate administration deviated negatively from calibrated values. However, low sample sizes (n=2/group, respectively) did not allow for statistical group comparisons.

**Supplemental Digital Content:**

**SDC 1. CwEPR difference spectra**

Figure SDC_1

Figure SDC_2

**SDC 2. P ENDOR simulation**

Figure SDC_3

Table SDC_1

**SDC 3. 7 Tesla MRI calibration in vitro**

Table SDC_2

Figure SDC_4

Table SDC_3

**SDC 4. CwEPR and $^1$H ENDOR spectra of contrast agents**

Figure SDC_5

Figure SDC_6

**SDC 5. Mn$^{2+}$ suppression on P ENDOR spectra**

Figure SDC_7

# Supplemental Digital Content 1 - CwEPR difference spectra

In Figure SDC_1, the subtraction of the control spectrum is shown on the example of the cortex biopsy from the EAE mouse treated with gadopentetate. As control, the cortex biopsy from the healthy, untreated mouse was used. A decomposition of the difference spectrum into the pure gadopentetate and free Gd spectra is shown in Figure SDC_2. The sum of the two components (gadopentetate: free $Gd^{3+}$ ratio 7:3) (blue) fits well the experimental difference spectrum (red).

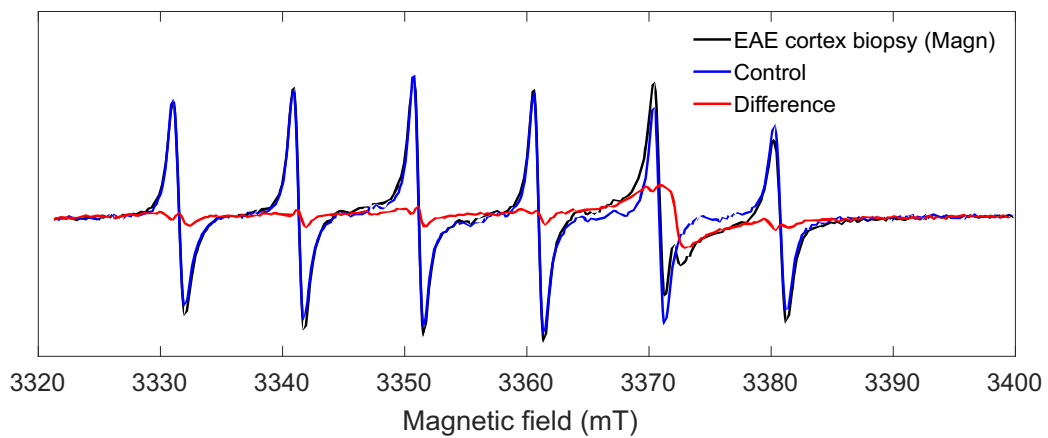

**Figure SDC_1.** CwEPR spectra of a cerebellar cortex biopsy from a *gadopentetate*-treated EAE mouse (black) in comparison to a cerebellar cortex biopsy of a HC untreated mouse (blue) and the difference (red).

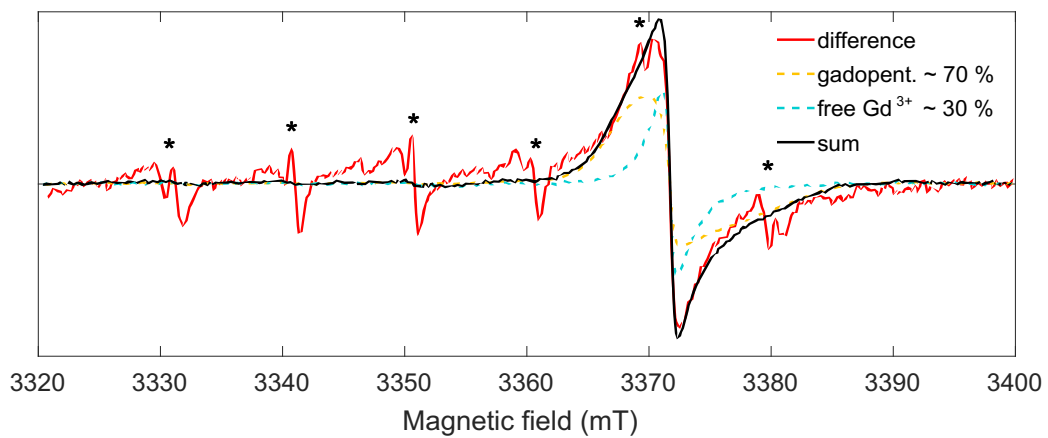

**Figure SDC_2**. Decomposition of the difference spectrum from Figure SDC_1 (red) into a spectrum of pure gadopentetate (dashed yellow) and of free $Gd^{3+}$ (dashed light blue). Residuals from the $Mn^{2+}$ contribution are marked with asterisks.

## Supplemental Digital Content 2 - P ENDOR simulation

To extract the Gd-P distance from the P ENDOR spectra, fits were performed using EasySpin's least-square fitting (esfit) and the pulsed EPR simulation function saffron (62). An electronic spin S=1/2 with g=1.99 was used as an approximation, since other transitions than the central transition ($m_S = -1/2 \leftrightarrow m_S = +1/2$) can be neglected at W-band due to broadening. As the exact coordination of Gd in these Gd-P systems was not known, only one P with axial hyperfine coupling was assumed. The predefined sequence "MimsENDOR" was used with an approximated pulse excitation width of 1/(18 ns). As fitting options, a baseline offset and least-square autoscaling were chosen, other settings were taken just as in the experiment. The hyperfine coupling A and the ENDOR linewidth were varied in the fit, and the distance $r_{Gd-P}$ between Gd and P (see Table SDC_1) could then be calculated from the dipolar part of the hyperfine interaction. The respective uncertainties, also given in Table SDC_1, were determined from error propagation of the standard deviations obtained from the fit. As lower limits, the RF frequency resolution of 0.01 MHz (one asterisk, for frequencies) as well as a spatial resolution of 0.01 nm (two asterisks, for distances) were employed if the error from the fit was unrealistically small, which represented our insufficient knowledge about the composition of the system.

In Figure SDC_3, the smoothed P ENDOR spectra (black) of the gadopentetate-treated EAE mouse, the references Gd-DMPC and Gd-ATP and the gadopentetate-treated slices with their respective fit results (red) are shown. A characteristic property of Mims ENDOR spectra are the τ-dependent so-called Mims holes, periodically occurring frequencies at which no signal can be detected. To prove that the choice of τ did not conceal crucial parts of the ENDOR spectrum, a Davies ENDOR spectrum of Gd-ATP, which does not contain any holes except at the center frequency, is also shown in gray. The spectrum closely resembles the one presented in a recent study of phosphate binding to lanthanide complexes (65).

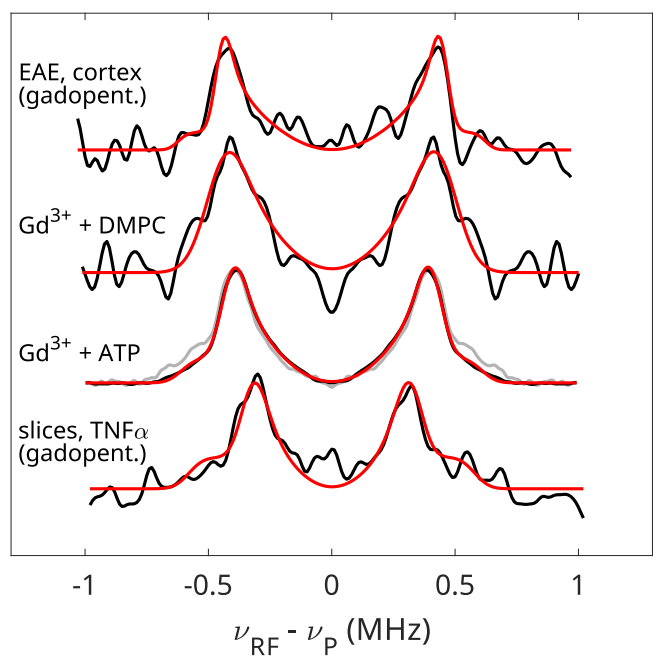

**Figure SDC_3.** P ENDOR spectra (black) and corresponding fits (red), performed using the EasySpin (62) function saffron. Gd-ATP was additionally measured with the Davies sequence (grey).

**Table SDC_1:** P ENDOR fit results. $A_{iso}$ and r were calculated from the axial hyperfine tensor, and uncertainties were determined from error propagation of the standard deviations. The lower limit for the error in frequency was derived from the frequency resolution (values marked with asterisks), and for the error in distance to represent the insufficient knowledge about the composition of the system (two asterisks).

|  | $A_{iso}$ (MHz) | $\Delta A_{iso}$ (MHz) | r (nm) | $\Delta$r (nm) | lwEndor (MHz) | $\Delta$lwEndor (MHz) |
|---|---|---|---|---|---|---|
| **EAE, cortex (gadopent.)** | -0.19 | 0.04 | 0.35 | 0.01 | 0.07 | 0.02 |
| **Gd-DMPC** | -0.3 | 0.2 | 0.36 | 0.03 | 0.17 | 0.05 |
| **Gd-ATP** | -0.15 | 0.01* | 0.36 | 0.01** | 0.11 | 0.01* |
| **slices, TNFα (gadopent.)** | -0.06 | 0.02 | 0.37 | 0.01** | 0.11 | 0.02 |

## Supplemental Digital Content 3 - 7 Tesla MRI calibration in vitro

As shown in Figure SDC_4 a) to _c), we measured the relaxivity $r_1$ of ascending GBCA standards in homogenized mouse brain tissue by linear fitting of $R_1$ in relation to concentrations of Gd (20 µM – 100 µM). The central concept is that $r_1$ is determined by the specific interaction of paramagnetic Gd with $^1$H. Higher $r_1$ values indicate a more efficient GBCA in shortening $T_1$ (33). Validation experiments demonstrated that both GBCAs exhibit similar $r_1$ values in water, in agreement with ranges previously reported in the literature (32, 36, 78). As detailed in Table SDC_2, $r_1$ increased for both gadopentetate and gadobutrol in FBS as well as in homogenized mouse brain compared to mean $r_1$ values in water, albeit to different extents, likely due to higher solution viscosity (79) and GBCA binding to endogenous macromolecules (33, 36, 78, 79). The computation of $T_1$ deviation for the medulla, cerebellar cortex, and CN 10 days post-injection of both tested GBCAs was based on quantification of Gd retention using previously acquired *in vivo* $T_1$ relaxometry and LA-ICP-MS data. While these underlying data were already available, only the Gd quantification for the CN had been published before (47); the present analysis newly includes the medulla and cortex (Figures SDC_4d, e, g, h). For better traceability of the T1 deviation computation, the CN data at day 10 post-GBCA application are shown again here, as illustrated in Figures SDC_4f and SDC_4i.

More specifically, the following derived average $R_1(0)$ values (baseline pre-GBCA administration) for the respective brain regions were used for the calculation of $T_1$ deviation: medulla (EAE vs. HC; 0.50 s$^{-1}$ vs 0.52 s$^{-1}$), cerebellar cortex (0.44 s$^{-1}$ vs 0.46 s$^{-1}$), CN (0.49 s$^{-1}$ vs 0.51 s$^{-1}$). As shown in Figure 1a, this pre-GBCA administration baseline corresponded to the EAE peak, where EAE mice showed the maximum clinical disease severity. 10 days post GBCA administration, average absolute $T_{1,iv}$ values of *n*=8-10 mice/experimental group were derived for the medulla (mean ± SD; EAE vs. HC; gadopentetate, 1.82 ± 0.05 vs 1.84 ± 0.10 s; gadobutrol, 1.82 ± 0.12 vs 1.87 ± 0.08 s), the cerebellar cortex (EAE vs. HC; gadopentetate, 1.98 ± 0.07 vs 2.10 ± 0.06 s; gadobutrol, 2.09 ± 0.04 vs 2.09 ± 0.06 s), and the CN (EAE vs. HC; gadopentetate, 1.81 ± 0.13 vs 1.90 ± 0.08 s; gadobutrol, 1.92 ± 0.11 vs 1.92 ± 0.10 s), based on the *in vivo* $T_1$ relaxometry data acquired previously (47). Percentage $T_1$ change was further computed for the three brain regions, to account for potential pre-contrast $T_1$ differences and visualize the effect of GBCA administration, as illustrated in Figure SDC_4d-f. Comparing EAE and HC mice, we saw significant differences in $T_1$ change 10 days post-gadopentetate administration for the medulla (adjusted p=0.029) and the cerebellar cortex (adjusted p=0.002), as well as 10 days post-gadobutrol for the medulla (adjusted p=0.037). After Bonferroni correction, we saw no significant differences between EAE and HC mice within the CN (post-gadopentetate, unadjusted p=0.046). For comparison, Gd concentrations in the CN (47), medulla and cerebellar cortex of EAE and HC mice 10 days after GBCA injection were determined based on the quantitative LA-ICP-MS images for Gd (Figure SDC_4).

**Table SDC_2:** Computation of $r_1$ of both GBCAs using 7 Tesla *in vitro* MRI of gadopentetate and gadobutrol standards in water, FBS, and homogenized mouse brain.

| $C_{Gd}$ (µM) | Gadopentetate | | | Gadobutrol | | |
|---|---|---|---|---|---|---|
| | $T_1$ (s) | $R_1$ (s$^{-1}$) | $r_1$ (mM$^{-1}$ s$^{-1}$) | $T_1$ (s) | $R_1$ (s$^{-1}$) | $r_1$ (mM$^{-1}$ s$^{-1}$) |
| *In water:* 0 | 4.00 | 0.25 | | 4.00 | 0.25 | |
| 20 | 3.31 | 0.30 | | 3.43 | 0.29 | |
| 40 | 2.49 | 0.40 | **3.37** | 2.52 | 0.40 | **3.22** |
| 60 | 2.24 | 0.45 | | 2.29 | 0.44 | |
| 80 | 2.03 | 0.49 | | 2.11 | 0.47 | |
| 100 | 1.67 | 0.60 | | 1.71 | 0.58 | |
| *In FBS:* 0 | 2.93 | 0.34 | | 2.88 | 0.35 | |
| 20 | 2.55 | 0.39 | | 2.48 | 0.40 | |
| 40 | 1.97 | 0.51 | **3.50** | 1.87 | 0.53 | **4.12** |
| 60 | 1.82 | 0.55 | | 1.69 | 0.59 | |
| 80 | 1.69 | 0.60 | | 1.58 | 0.63 | |
| 100 | 1.42 | 0.70 | | 1.29 | 0.77 | |
| *In brain:* 0 | 1.97 | 0.51 | | 2.01 | 0.50 | |
| 20 | 1.82 | 0.55 | | 1.60 | 0.62 | |
| 40 | 1.48 | 0.68 | **4.35** | 1.22 | 0.82 | **6.96** |
| 60 | 1.33 | 0.75 | | 1.09 | 0.92 | |
| 80 | 1.25 | 0.80 | | 0.99 | 1.01 | |
| 100 | 1.05 | 0.95 | | 0.82 | 1.22 | |

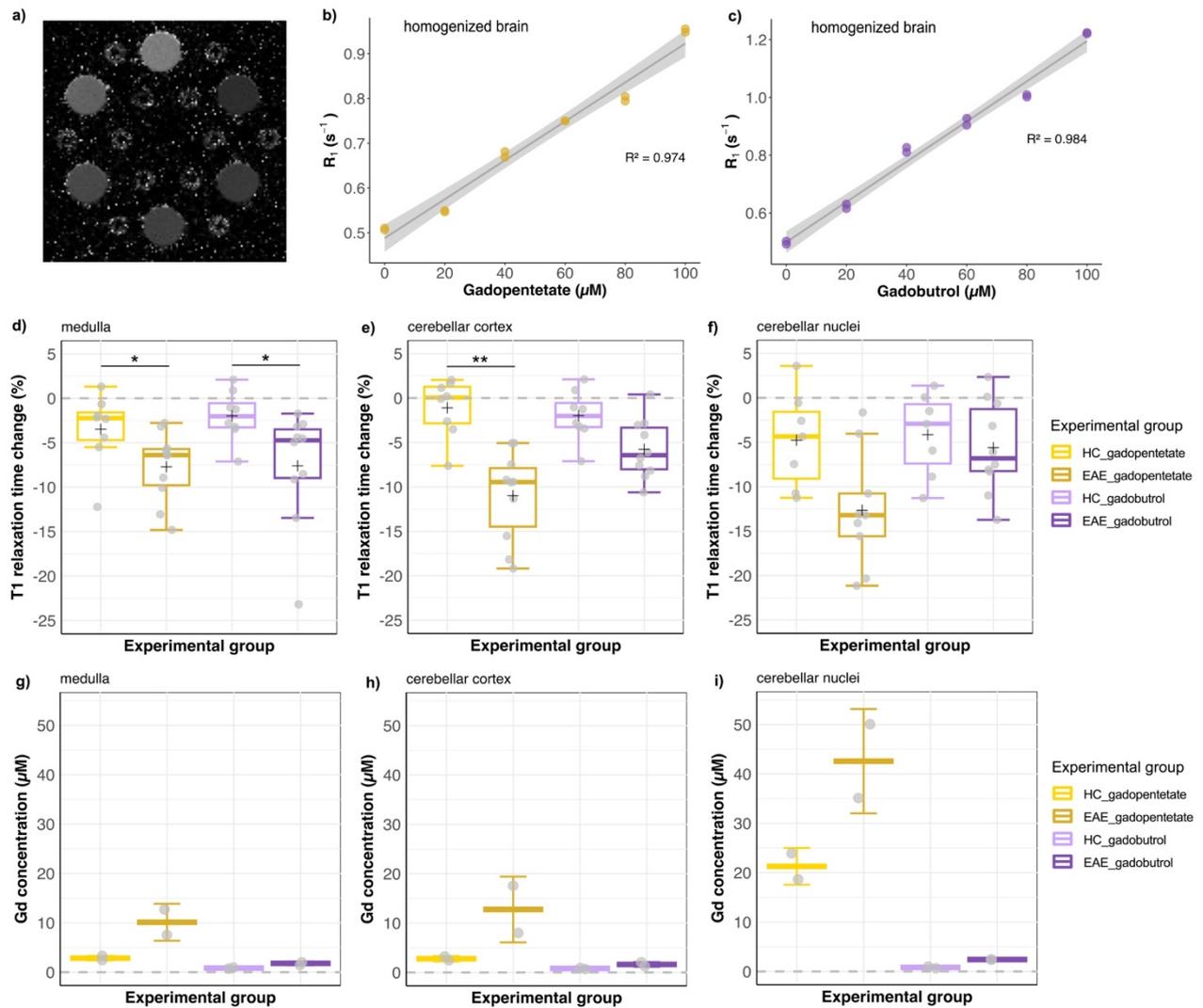

**Figure SDC_4. a)** Exemplary $T_1$ map image of simultaneously scanned NMR calibration tubes with increasing concentrations (20-100 μM, clockwise ascending) of gadobutrol. **b), c)** Determination of the $T_1$-relaxivity ($r_1$) by linear fitting of longitudinal relaxation rates ($R_1=T_1^{-1}$) in relation to increasing concentrations of gadopentetate and gadobutrol. **d), e), f)** $T_1$ relaxation time change (in %) of the medulla, cerebellar cortex, and CN of EAE and HC mice 10 days post-injection of both tested GBCAs (n=8-10/group). Tukey boxplots: lines represent the median, + indicates the mean, boxes span the interquartile range (IQR), whiskers extend to 1.5 × IQR. Mann-Whitney tests with Bonferroni correction across two tested GBCAs, with * implying $p < 0.05$ and ** implying $p < 0.01$. **g), h), i)** LA-ICP-MS quantification of Gd concentrations within the medulla, cerebellar cortex, and CN of EAE and HC mice 10 days post-injection. Low sample sizes for LA-ICP-MS (n=2/group, respectively) (47) did not allow for statistical group comparisons. Lines represent the mean; whiskers indicate ± 1 SD.

**Table SDC_3:** Results of semi-quantitative analysis of Gd retention in cerebella of EAE or HC mice after treatment with GBCAs detected by LA-ICP-MS.

| $C_{Gd}$ (µM) | | HC Gadopentetate | | EAE Gadopentetate | | HC Gadobutrol | | EAE Gadobutrol | |
|---|---|---|---|---|---|---|---|---|---|
| | | | *mean ± SD* | | *mean ± SD* | | *mean ± SD* | | *mean ± SD* |
| cerebellar nuclei | #1 | 23.9 | 21.3 ± 3.7 | **50.1** | 42.6 ± 10.6 | 1.0 | 0.8 ± 0.2 | 2.4 | 2.4 ± <0.1 |
| | #2 | 18.7 | | 35.1 | | **0.7** | | **2.4** | |
| cerebellar cortex | #1 | 3.1 | 2.8 ± 0.5 | **17.5** | 12.8 ± 6.7 | 1.0 | 0.8 ± 0.2 | 2.0 | 1.6 ± 0.5 |
| | #2 | 2.4 | | 8.0 | | **0.6** | | **1.3** | |
| medulla | #1 | 3.2 | 2.8 ± 0.5 | **12.8** | 10.1 ± 3.7 | 1.1 | 0.8 ± 0.3 | 2.1 | 1.8 ± 0.4 |
| | #2 | 2.5 | | 7.5 | | **0.6** | | **1.5** | |

# Supplemental Digital Content 4 - CwEPR and ¹H ENDOR spectra of contrast agents

Figure SDC_5 shows the comparison of the cwEPR spectra of gadopentetate (yellow), gadobutrol (light blue), and $GdCl_3$ (purple). While gadopentetate has a distinct shape, gadobutrol and $GdCl_3$ are very similar. However, with ¹H ENDOR, they are easily distinguishable (Figure SDC_6).

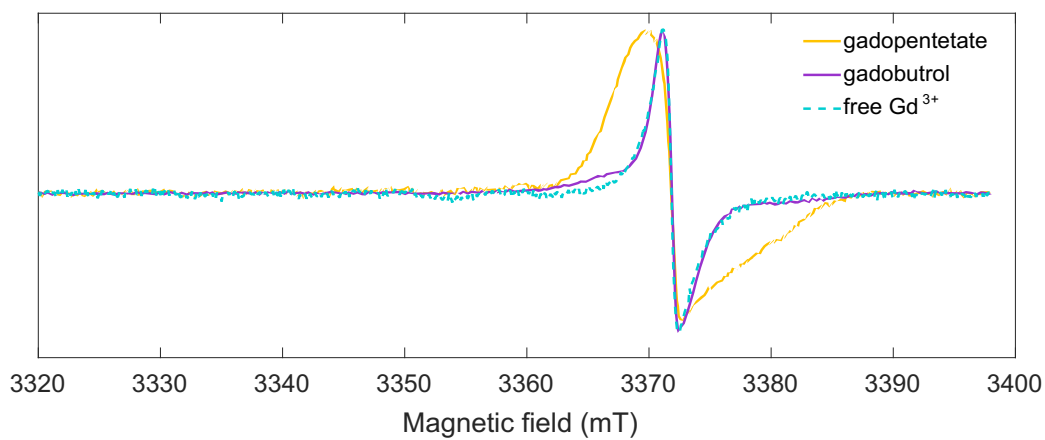

**Figure SDC_5.** CwEPR spectra of gadopentetate, gadobutrol and $GdCl_3$.

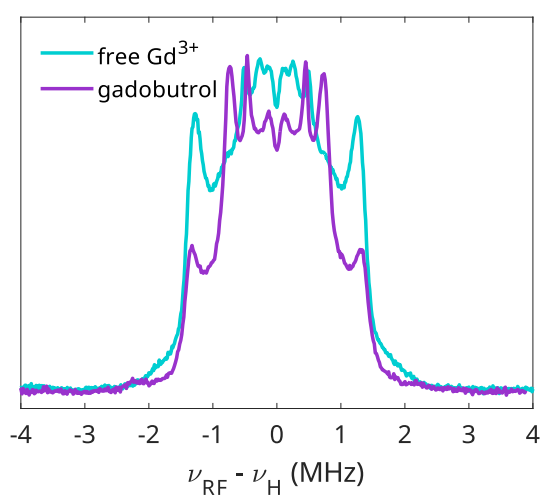

**Figure SDC_6.** ¹H ENDOR spectra of gadobutrol and $GdCl_3$.

# Supplemental Digital Content 5 - Mn$^{2+}$ suppression on P ENDOR spectra

Figure SDC_7 shows the P ENDOR spectra of the gadopentetate-treated EAE mouse measured on Gd with the Mn suppression sequence (black) in comparison to the spectrum measured on the lowest-field Mn$^{2+}$ line with the sequence optimized for Mn (blue). This spectrum contains only purely resolved intensity in the –0.2 to +0.2 MHz frequency range, which shows the absence of P in the direct environment of Mn. In the spectrum measured on Gd, however, no intensity in this range is visible, confirming the efficiency of the Mn suppression.

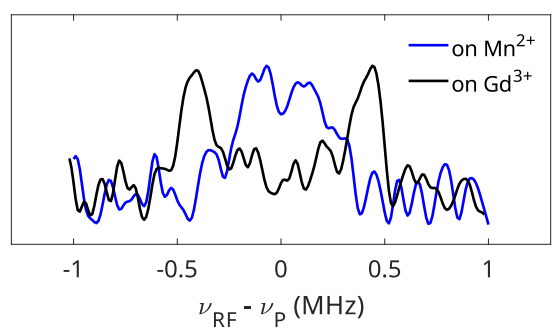

**Figure SDC_7.** P ENDOR spectra of the gadopentetate-treated EAE mouse measured on Mn$^{2+}$ (blue) and Gd with Mn suppression (black).